\documentclass[a4paper,aps,prl,superscriptaddress,floatfix,nofootinbib,twocolumn,longbibliography]{revtex4-1}

\usepackage{bbold}
\usepackage{bbm}
\usepackage[pdftex]{graphicx}
\usepackage{latexsym,amsmath,verbatim,amssymb,txfonts}
\usepackage{color}
\usepackage{rotating}
\usepackage{verbatim}
\usepackage{multirow}
\usepackage[english]{babel}
\usepackage{comment}
\usepackage{hyperref}
\hypersetup{colorlinks=true, citecolor=blue, linkcolor=blue, urlcolor=black}

\begin{document}

\title{Characterizing the analogy between hyperbolic embedding and
  community structure of complex networks}

\author{Ali Faqeeh}
\affiliation{MACSI, Department of Mathematics and Statistics, University of Limerick, Limerick, Ireland}
\affiliation{Center for Complex Networks and Systems Research, School
  of Informatics, Computing, and Engineering, Indiana University, Bloomington,
  Indiana 47408, USA}

\author{Saeed Osat}
\affiliation{Quantum Complexity Science Initiative, Skolkovo Institute
  of Science and Technology,
Skoltech Building 3,
Moscow, 143026, Russia}

\author{Filippo Radicchi}
\affiliation{Center for Complex Networks and Systems Research, School
  of Informatics, Computing, and Engineering, Indiana University, Bloomington,
  Indiana 47408, USA}
\email{filiradi@indiana.edu}

\begin{abstract}

We show that the community structure of a network can be used as a
coarse version of its embedding in a hidden space with
hyperbolic geometry.
The finding emerges from a systematic analysis
of several real-world and synthetic networks.
We take advantage of the analogy
for reinterpreting results originally
obtained through network hyperbolic embedding
in terms of community structure only. First,
we show that the robustness of a multiplex
network can be controlled
by tuning the correlation between the community structures
across different layers. Second, we deploy an efficient
greedy protocol for network navigability that makes
use of routing tables based on community structure.
\end{abstract}

\maketitle

%%Introduction

%%Hyperbolic mapping
A wealth of recent publications provides
evidence of the advantages that may arise from thinking of
real-world networks as instances of random network models
embedded in hidden metric
spaces~\cite{serrano2008self, boguna2009navigability}. In this class of models,
every node is represented by coordinates that identify its
position in the underlying space, and the distance between
pairs of nodes determines their likelihood of being connected.
The most popular formulation
of spatially embedded network models
relies on  hyperbolic geometry~\cite{krioukov2009curvature,
  krioukov2010hyperbolic}.
Hyperbolic network geometry emerges
spontaneously from
models of growing simplicial complexes~\cite{bianconi2017emergent}.
Hyperbolic geometry appears the natural choice for
networks with broad degree
distributions, under the hypothesis
that the generating mechanism for edges in the network
is a compromise between popularity of individual nodes
and similarity among pairs of
nodes~\cite{papadopoulos2012popularity}.
 Popularity is represented by the radial coordinate
 of nodes in the hyperbolic space,
 while similarity is accounted for by
the difference between angular
coordinates of pairs of nodes. Hyperbolic maps are useful in practical contexts, as
generating efficient routing protocols in information
networks~\cite{boguna2010sustaining}, characterizing hierarchical organization of
biochemical pathways in cellular
networks~\cite{serrano2012uncovering}, and monitoring the evolution of the international trade
network~\cite{garcia2016hidden}. However, thinking of networks
as embedded in the hyperbolic space is important
from the theoretical point of view too.
 Growing network models that rely on
hyperbolic geometry provide a genuine explanation
for the emergence of power-law degree distributions from local optimization
principles only~\cite{papadopoulos2012popularity}. Further,
recent work show that the main features of the percolation transition
in multiplex networks can be predicted by simply accounting
for inter-layer correlation among hyperbolic coordinates of
nodes~\cite{kleineberg2016hidden, kleineberg2017geometric}.

%%Stochastic Block Model and Community detection
Popularity and similarity are core features of models
used in network hyperbolic embedding. They are, however,
central in another heavily used
model in network science: the degree-corrected stochastic block
model (SBM)~\cite{karrer2011stochastic}.
The SBM assumes a hidden cluster
structure where nodes are divided into a
certain number of groups.
This classification
accounts for similarity, as pairs of nodes
have different likelihoods of being connected depending
on their group memberships. The degree correction provides instead
a natural way of accounting for the popularity of the individual nodes.
The SBM is generally considered in the context of graph clustering,
representing a generative network model with built-in
mesoscopic structure~\cite{fortunato2010community}.
The SBM is used in the formulation of principled
community detection methods~\cite{peixoto2017bayesian}.
These methods, in turn, are equivalent to other
well-established techniques for community
detection, giving therefore to the SBM
a central role in the graph clustering
business~\cite{newman2016equivalence}.

%%Analogy between Hyperbolic Model and SBM
At least superficially, the analogy
between the ideas of hyperbolic embedding and
community structure is apparent. In a recent paper, Wang {\it et
  al.} showed that information about community structure
can be used to improve accuracy and efficiency of standard algorithms for
hyperbolic embedding~\cite{wang2016hyperbolic}. Also,
previous work was devoted to the development of network models
embedded in hyperbolic geometry with the addition of a
pre-imposed community  structure~\cite{zuev2015emergence,
  garcia2017soft, muscoloni2017nonuniform}.
We are not aware, however, of previous attempts to
investigate the theoretical and practical similarity of
the two approaches when applied independently to
the same network topology.
This is the purpose of the present paper.

%%%Results

%%%%%%%%%%%%%%%%%%%

We assume that
the topology of an undirected and unweighted network $G$ with $N$ nodes
is fully specified by its adjacency matrix $A$, whose
element $A_{i,j} = A_{j,i} = 1$ if a connection between nodes
$i$ and $j$ is present, or $A_{i,j} = A_{j,i} = 0$, otherwise.
The hyperbolic embedding of the network $G$ consists in a pair of
coordinates $(r_i, \theta_i)$ for every node $i \in G$.
The quantity $r_i$ is the radial
coordinate of node $i$;
$\theta_i$ is its angular coordinate. We assume that this information
is at our disposal. The way we acquire such a knowledge depends on
whether the network analyzed is synthetic or real.
For synthetic graphs, we consider single instances of the
popularity-similarity optimization model
(PSOM)~\cite{papadopoulos2012popularity}, so that hyperbolic
coordinates correspond to ground-truth values of the model.
 We analyze also several real networks, where coordinates of
nodes are obtained by fitting graphs
against the PSOM. In this second scenario,
we either rely on embeddings publicly
available~\cite{kleineberg2016hidden, papadopoulos2015network_b} or we
apply publicly available algorithms to the graphs
~\cite{papadopoulos2015network_b}. Details are provided in the
Supplemental Material (SM). We remark that the PSOM is the model
of reference in most of the hyperbolic embedding techniques.
It assumes the existence of an underlying hyperbolic space, and
consists in a random growing network model
where nodes are connected depending on their distance,
and the value of other model parameters, such as
average degree $\langle k \rangle$, exponent $\gamma$ of
the power-law degree distribution $P(k) \sim k^{-\gamma}$, and
temperature $T$.  When a real network is fitted against the PSOM,
the parameters $\langle k \rangle$ and $\gamma$ of the model are
determined on the basis of the observed network, while
$T$ is treated as a free parameter~\cite{papadopoulos2015network_b}.
Its value may be set to the one that yields the best match between
theoretical and numerical results for the distance
dependent connection probability~\cite{papadopoulos2015network};
when hyperbolic embedding is used in greedy routing, one may look for
the $T$ value  that results in the highest success rate~\cite{papadopoulos2015network_b}.
The radial coordinate $r_i$ of every node $i$ is uniquely identified
by its degree $k_i$, hence $r_i$ doesn't require to be truly
learned. The angular coordinate $\theta_i$ for every node
$i \in G$ is instead  treated as a fitting parameter.
There are various techniques to perform the fit,
including approximated optimization algorithms~\cite{papadopoulos2015network,
papadopoulos2015network_b},
and {\it ad-hoc} heuristic methods~\cite{alanis2016efficient,
  muscoloni2017machine}.

In our analysis,  we further assume to know the
community structure of the graph $G$, consisting in a flat partition 
of the network into
$C$ total communities, where every node $i \in G$ is associated with a
discrete-valued coordinate $\sigma_i = 1, \ldots, C$.
Algorithms for community detection are
numerous~\cite{fortunato2010community}.
Here, we rely on results obtained by three popular methods:
the Louvain algorithm~\cite{blondel2008fast},
Infomap~\cite{rosvall2008maps}, and the algorithm by Ronhovde and
Nussinov~\cite{Ronhovde_pre10}. We remark that,
in the degree-corrected SBM, the probability for nodes $i$ and $j$ to
be connected is a function of $\sigma_i$, $\sigma_j$, $k_i$ and $k_j$.
Hence, the graph $G$ can be thought as embedded into
a community structure, where every node $i$ is
{\it de facto} represented by the
coordinates $(k_i, \sigma_i)$.

\begin{figure}[!tb]
\hspace*{-.62cm}
\includegraphics[width=0.52\textwidth]{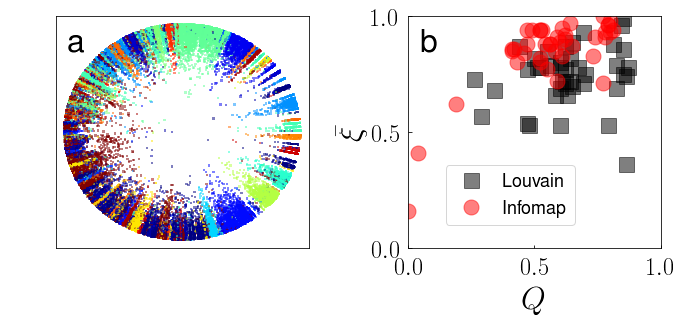}

\caption{Hyperbolic embedding and community structure for real and
  synthetic networks.
(a) We compare the hyperbolic embedding of the IPv4
Internet
with its community structure.  Every point represents a node in the largest
connected component of the graph.
Positions are
determined by the radial and angular coordinates of the nodes in the
hyperbolic embedding of the network~\cite{kleineberg2016hidden}.
We use the best partition found by the Louvain algorithm to determine the
community structure of the graph~\cite{blondel2008fast}. The partition consists of $C=31$
communities. Colors of the points identify community
memberships.  The value of the modularity is $Q = 0.61$, while
angular coherence is $\bar{\xi} = 0.72$.
 (b) We consider $39$ real-world networks and $2$ instances of the PSOM, and
compare their community structure and hyperbolic embedding (see
details in~SM). The plot displays each network on the
$(Q,\bar{\xi})$-plane. 
We show results obtained using Louvain
(black squares) and Infomap (red circles)~\cite{rosvall2008maps}.
}
\label{fig:1}
\end{figure}

A direct comparison between the hyperbolic embedding and the
community structure of the graph $G$ consists in a comparison
between the coordinates of the individual nodes in the two
representations. Further, as the
degree of the nodes trivially matches in both representations, the
comparison reduces only in matching angular coordinates
$\theta$s and group memberships $\sigma$s.
From the numerous empirical tests we conducted on both real and
synthetic networks, two main conclusions emerge. First, networks
usually considered in hyperbolic embedding applications are highly
modular, in the sense that partitions found by community detection
algorithms correspond to very large values of the modularity function
$Q$~\cite{newman2004finding} (see Figure~\ref{fig:1} and
~SM.  Second,  nodes  within the same communities are
likely to have similar angular coordinates.
We note that this second finding is in line
with what already shown in Ref.~\cite{wang2016hyperbolic}.
To quantify coherence
among angular coordinates of nodes within the
same community $g$,  we first
define the variables $\xi_g$ and
$\phi_g$ with
\begin{equation}
\xi_g \; e^{\textrm{i} \, \phi_g}  = \frac{1}{n_g} \, \sum_{j=1}^N
\delta_{\sigma_j, g} \, e^{\textrm{i} \,
 \theta_j} \; .
\label{eq:comm_angular}
\end{equation}
$\delta_{x,y} = 1$ if $x=y$ and $\delta_{x,y} = 0$, otherwise.
The r.h.s. of
Eq.~(\ref{eq:comm_angular}) stands for the
sums of vectors in the complex plane
of the type $e^{\textrm{i} \,  \theta} =
\cos(\theta) +  \textrm{i} \sin(\theta) $ of all nodes in
group $g$.
The vectorial sum is divided by the community size $n_g$
to obtain an average vector for the community.
$\phi_g$ is the angular coordinate
of community $g$.  The module $0 \leq \xi_g \leq 1$ indicates how coherent
are the angular coordinates of the nodes within group $g$. Note that
the definition of Eq.~(\ref{eq:comm_angular}) resembles
the one used for the order parameter of the Kuramoto
model~\cite{kuramoto2012chemical}. We finally measure the
angular coherence of a partition as the
weighted average
\begin{equation}
\bar{\xi} = \frac{1}{N} \, \sum_{g=1}^C  n_g \xi_g    \; .
\label{eq:mean_angular}
\end{equation}
By definition, we have that $0 \leq \bar{\xi} \leq 1$.
For all networks considered in our
analysis (see Figure~\ref{fig:1} and
~SM), angular coherence is typically large.

Our empirical tests demonstrate that
strong angular coherence within communities
of strongly modular
networks is a quite robust feature of both synthetic and
real systems.  This finding tells us that the analogy
between community structure and hyperbolic embedding may extend
beyond the mere similarity among their ingredients.
The following examples
show that the analogy is useful
also in the interpretation of physical properties of networks and
the design of practical algorithms on networks.

%%%%%%%%
%%%%%%%%%

\begin{figure}[!t]
\begin{center}
\includegraphics[width=0.45\textwidth]{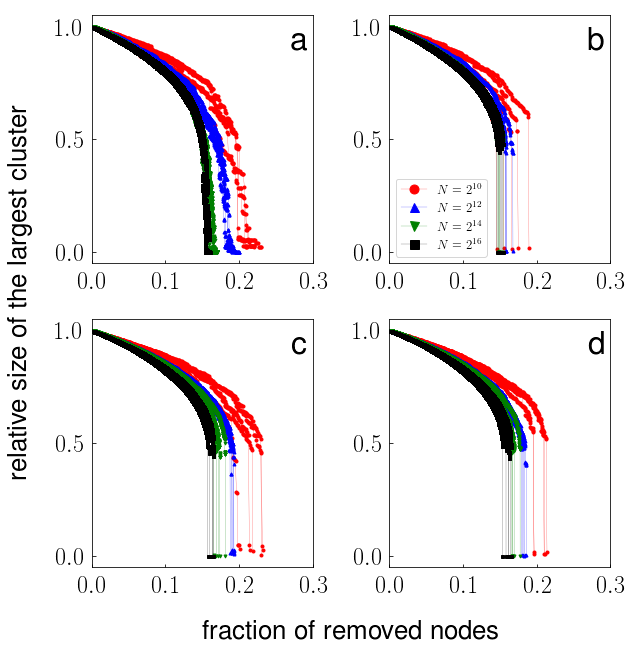}
\end{center}
\caption{Robustness of multiplex networks with correlated community
  structure. We measure the relative size of the largest mutually
  connected cluster
  as a function of the fraction of nodes removed
  from the system. The
  synthetic
  multiplex graphs
  are
  obtained using the recipe described in the text,  where two
Lancichinetti-Fortunato-Radicchi (LFR)
networks
with
 size $N$ are coupled together.
 The LFR models are such that: the average degree is $\langle k \rangle = 6$;
 the maximum degree is $k_{max} = \sqrt{N}$; node degrees $k$ obey a power-law
 distribution $P(k) \sim k^{-\gamma}$ with exponent $\gamma = 2.6$;
 there are $C = \sqrt{N}$
 communities of identical size $S = \sqrt{N}$.
 For every
 $N$,  we show the results for five distinct
 instances of the model. (a) LFR graphs are generated with
 $\mu = 0.1$. Labels are exchanged only among
 nodes within the same clusters. All nodes are considered for
 relabeling at least once. (b) Same as in panel a. However,
 relabeling of nodes is not constrained by community
 structure. The number of nodes that are relabeled is such that
 the edge overlap among layers is the same as in panel a~(SM). (c and d) Same as in panels a and b, respectively, but for
LFR graphs constructed using $\mu = 0.3$.
}
\label{fig:2}
\end{figure}

%%%%%%%%%%%%%
Our first example regards the rephrasing,
in terms of community structure only, of a
result obtained by analyzing the hyperbolic
embedding of multiplex networks.
In two recent papers~\cite{kleineberg2016hidden,
  kleineberg2017geometric},  Kleineberg and collaborators
found that inter-layer correlation between
hyperbolic coordinates of nodes in multiplex networks
is a good predictor for the
robustness of a system under targeted attack.
Specifically, they found that, when correlation among angular
coordinates is high,
the percolation transition is smooth. Instead,
multiplex networks
characterized by a small value of inter-layer correlation
exhibit abrupt percolation transitions. The finding was
initially obtained for real-world multiplex networks.
A theoretical explanation
was then given in terms of a synthetic network
model~\cite{kleineberg2017geometric}.
To further support the analogy between hyperbolic embedding and
community structure that we are arguing for in this paper,
we replicated all results of
Ref.~\cite{kleineberg2017geometric}
using community structure only.
First, we analyzed the same real-world multiplex networks considered
in Ref.~\cite{kleineberg2017geometric}. We found that their
robustness can be predicted very well
by the level of correlation
among the community structures of the layers (see SM).
Then, we provided a theoretical explanation.
We replaced the network model by Kleineberg {\it et al.}
with a variant of the SBM known in the literature as
the
Lancichinetti-Fortunato-Radicchi (LFR) benchmark
graph~\cite{lancichinetti2008benchmark}.
The LFR model mostly differs from the standard SBM for relying on
heterogeneous distributions of node degrees and community sizes.
In our model for multiplex networks (see SM),
we first generate a single LFR graph that is used as
the topology for both layers.
We then exchange the node labels in one layer
to destroy edge overlap and degree-degree correlation.
We consider two distinct
scenarios. In the first case,  we exchange the
label of every node with the one of a randomly chosen node
from the same community. This allows us to maintain
perfect correlation between the community structure of the two layers.
In the second case, we exchange the labels of a number of randomly
sampled nodes such that the edge overlap between the layers equals the value
obtained in the first randomization scheme. This second recipe
completely destroys correlation between the community structures
of the two layers.
In Fig.~\ref{fig:2}a, we show the phase diagrams for instances of the
multiplex model when
relabeling uses information about the community structure of the
graph. Here, the community structure is strong, in the sense that
the fraction of external connections per node is only $\mu = 0.1$.
The transition appears smooth, and becomes smoother as the size of the
model increases. This is an indication that, in the limit of
infinitely large LFR graphs, the percolation transition
is likely continuous. In Fig.~\ref{fig:2}b, we consider the
second relabeling scheme that doesn't account for community
structure. The resulting diagrams indicate that the percolation
transition is abrupt. The level of
correlation among community structure of the two layers can be decreased
by increasing $\mu$, so that community structure itself becomes less
neat. This is done in Figs.~\ref{fig:2}c and d, where the transition appear
abrupt no matter how the labels of the nodes are relabeled. In ~SM, we
report results for different parameter values of the LFR model. Results
confirm our claim that the extent of correlation between the community
structure of the layers of a multiplex can be used to explain
robustness properties of
the system under targeted attack.

%%%%%%%%
%%%%%%%%%

%%%%%%%%
\begin{figure}[!b]
\begin{center}
\includegraphics[width=0.49\textwidth]{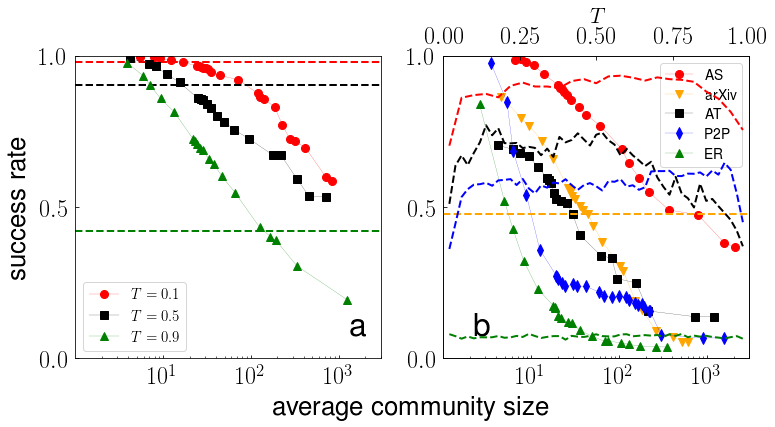}
\end{center}
\caption{Performance of community-based routing. (a) We consider single instances
of the growing network model of Ref.~\cite{papadopoulos2015network}
with $N = 5,000$ nodes, $\langle k
\rangle = 5$, and degree exponent $\gamma = 2.1$. Different symbols and
colors refer to different values of the temperature $T$.
The plot
shows how success rate of the community-based greedy routing
strategy changes as a function the average size of the
communities.
Communities are identified using the algorithm
by Ronhovde and Nussinov~\cite{Ronhovde_pre10}.
Their number
can be varied by changing the resolution level of the algorithm.
Dashed lines are obtained on the same networks but using
hyperbolic greedy routing. (b) Same as in panel a, but for
real networks. We consider the following networks: the
Internet at the level of autonomous systems
(AS)~\cite{leskovec2005graphs}; the worldwide air transportation
network (AT)~\cite{Airports};
the European road network  (ER)~\cite{Euro_roads};
the peer-to-peer network (P2P)~\cite{p2p_1};
the  arXiv collaboration network~\cite{de2015muxviz}.
For all the networks (except arXiv) the dashed lines are obtained by varying the temperature $T$ in the
algorithm for hyperbolic embedding introduced in
Ref.~\cite{papadopoulos2015network_b}; for the arXiv network the dashed line shows the result for the optimum hyperbolic coordinates whose data was available in \cite{kleineberg2016hidden}.
Details can be found in~SM.
}
\label{fig:3}
\end{figure}
%%%%%%%%%
Our second example focuses on
greedy routing~\cite{boguna2009navigability, boguna2010sustaining}.
To be brief, the scenario considered is the following. A packet
originated by node $s$ must be delivered to node $t$. The packet can
navigate the network by walking at each step on an edge.
The packet moves on the network till it reaches its destination $t$, or
it visits twice the same node. In the first case, the packet is
correctly delivered. In the second case, the packet is considered
lost, and it is discarded.
The goal of a good routing strategy is to deliver packets with high
probability and with a small number of steps, for
any randomly chosen pair of source and target nodes $s$ and $t$.
Hyperbolic embedding turns out to be very useful in the formulation of
a greedy strategy, where individual steps
are determined on the basis of the distance among nodes in the
hyperbolic space. Specifically, if a message  is at node $i$, then the
next move will be on the node
\begin{equation}
j^{\,(i)}_{(best)} = \arg\,  \min_{j \in \mathcal{N}_i }  d(j, t) \; ,
\label{eq:routing}
\end{equation}
where
$\mathcal{N}_i$ is the set of neighbors of $i$, and $d(j, t)$
is the
distance between nodes $j$ and $t$.
The greedy technique is computationally feasible as every node
needs to know only the identity and the geometric coordinates
of its neighbors. The regimes
of effectiveness of the routing method have been systematically
studied in artificial network models~\cite{boguna2009navigability}.
The technique has been proven to be
extremely effective on some real-world topologies~\cite{boguna2009navigability, boguna2010sustaining}.
We devised a new greedy routing protocol that makes use of
the cluster structure of a network instead of its hyperbolic
embedding.  Specifically, we replaced the
definition of distance in the hyperbolic space between nodes
with the fitness
\begin{equation}
d(j,t) = \beta D_{\sigma_j, \sigma_t} - (1 - \beta) \ln { k_j } \; ,
\label{eq:distance}
\end{equation}
where $k_{j}$ is the degree of node $j$, and
$\sigma_j$ and $\sigma_t$ are
the indices of the communities of
nodes  $j$ and $t$, respectively.
$D_{\sigma_j, \sigma_t}$ is the length of the shortest path between
communities $\sigma_j$ and $\sigma_t$ calculated on a weighted
supernetwork in which supernodes are communities of the original
network. Each pair of supernodes $g$ and $q$ is connected with a
superedge with weight $1-\ln {\rho_{g,q}}$; here $\rho_{g,q}$ is the
probability that, in the original network, a randomly chosen node in
community $g$ has an edge to community $q$~(see SM).
The term $\ln { k_j }$ in Eq.~(\ref{eq:distance}) serves
to perform degree correction.
The factor $0 \leq \beta \leq 1$ serves to control
the relative importance of one factor over the other.
$\beta$ plays a similar role as of the
temperature $T$ in hyperbolic routing
protocols~\cite{papadopoulos2015network_b}, and its value may be
appropriately chosen with the goal of optimizing the success rate in
the delivery of messages~(see SM).
The routing protocol based on Eq.~(\ref{eq:distance})
is still computationally efficient as long as the total
number of communities $C$ grows sub-linearly with the
size of the graph $N$.
In the extreme case, where
every community is formed by a single node, so that $C = N$, the
method will be $100\%$ accurate in delivering packets, but also
computationally expensive.
In Figure~\ref{fig:3}, we display the performance of
community-based
greedy routing  as a function of the mean size of the communities.
We study the performance on both synthetic and real-world networks.
The number of communities is tuned by changing the resolution
parameter in the algorithm by Ronhovde and Nussinov~\cite{Ronhovde_pre10}.
Success rates of the community-based greedy protocol are always very
good, as long as communities are not too large.

%%Conclusions

In summary, we showed that looking at a network
as embedded in a hyperbolic geometry is similar, both in theory
and practice, to pretending that the network is
organized into communities, 
provided that community structure is detected by
a method that accounts for the degree of the nodes.
Our finding provides evidence that the inter-community
  structure in networks may have geometric organization, meaning that at
  the global level, geometry dominates, while at the local
  scale, community
  memberships prevail. Thus, real networks 
may be modeled
by a graphon~\cite{Lovasz12} consisting of a mixture of latent-spatial 
and block-like structures. 
This fundamental model has the potential to generate further
understanding of physical processes, such as 
spreading and 
synchronization, in real networks.

\begin{acknowledgements}
The authors thank G. Bianconi, C. V. Cannistraci, D. Krioukov , and M.\'A. Serrano for comments on the manuscript.
A.F. and F.R. acknowledge support from the U.S. Army Research Office (W911NF-16-1-0104).
F.R. acknowledges support from the National Science Foundation
(CMMI-1552487).
A.F. acknowledges support from the Science Foundation Ireland (16/IA/4470).
\end{acknowledgements}

%%%%%%%%%%%%
%%%%%%%%%%%%
%%%%%%%%%%%%

\clearpage
\newpage

%\begin{NoHyper}
%\setcounter{page}{1}
\renewcommand{\theequation}{S\arabic{equation}}
\setcounter{equation}{0}
\renewcommand{\thefigure}{S\arabic{figure}}
\setcounter{figure}{0}
\renewcommand{\thetable}{S\arabic{table}}
\setcounter{table}{0}

\renewcommand{\theHfigure}{Supplement.\thefigure}
\renewcommand{\theHequation}{Supplement.\theequation}
\renewcommand{\theHtable}{Supplement.\thetable}
\newcommand*{\citenolink}[1]{%
  \begin{NoHyper}\cite{#1}\end{NoHyper}%
}
\hypersetup{urlcolor=black}

\section{Supplemental Material}

\subsection{Hyperbolic embedding and community detection}

In table~\ref{table}, we provide a list of all networks considered in
our analysis.

We obtain hyperbolic coordinates of networks in the following way.
For real networks, we either rely on embeddings publicly
available~\cite{kleineberg2016hidden, papadopoulos2015network_b} or we
apply publicly available algorithms to the graphs
~\cite{papadopoulos2015network_b}.
Urls of electronic resources for all networks are provided in table~\ref{table}.
In the hyperbolic embeddings that we performed, we
made use of the algorithm provided in~\url{https://bitbucket.org/dk-lab/2015_code_hypermap}.
As prescribed in
Ref.~\cite{papadopoulos2015network_b},  the value of
the temperature $T$ used in the embedding corresponds to the one
leading to maximal success rate in greedy routing ~\cite{boguna2009navigability,
boguna2010sustaining} (see section below).
 We further consider two instances
of the   popularity-similarity optimization model
(PSOM)~\cite{papadopoulos2012popularity}.
They are generated using different values of the
model parameters. The code
to generate instances of the PSOM
has been taken from \url{https://www.cut.ac.cy/eecei/staff/f.papadopoulos}.

We use three distinct methods for detecting communities in networks:
the Louvain algorithm~\cite{blondel2008fast}, Infomap~\cite{rosvall2008maps}, and
the algorithm by Ronhovde and
Nussinov~\cite{Ronhovde_pre10}.
Louvain and Infomap are used in the analysis about the relation
between hyperbolic embedding and community structure (see
Table~\ref{table}). The algorithm by Ronhovde and
Nussinov is used in the analysis of greedy routing.
For Louvain and Infomap we rely on the
algorithms implemented in the library \url{http://igraph.org/python}.
We consider always the ``best''  (i.e., the one with maximum modularity for Louvain
, the one with minimum description
length for Infomap) partitions found by the algorithms.
The implementation of the algorithm by Ronhovde and
Nussinov was taken from~\url{http://www.elemartelot.org/index.php/programming/cd-code}.
We chose this algorithm to study greedy routing as it allows for a
finer tuning of the resolution of the community structure than the
other two algorithms. After obtaining the modular structure from this algorithm, we perform an additional step to improve the quality of communities: If there is any community with size one we change the community label of the only member of that community to the label of its highest degree neighbor.
%%%%%%%%%%%

\begin{table*}[t]
\caption{
Relation between community structure and hyperbolic embedding
in real and synthetic networks.
From left to right, we report: name of the network, size of
the giant component $N$, number of communities $C$ identified by the Louvain
algorithm, value of the modularity $Q$ corresponding to the  Louvain
partition,
angular coherence $\bar{\xi}$ of the Louvain partition,  number of
communities $C$ identified by Infomap,
value of the modularity $Q$ corresponding to the Infomap
partition,
angular coherence $\bar{\xi}$ of the Infomap partition, reference(s)
of the papers where the dataset was reported and/or where
hyperbolic coordinates of the network were obtained, urls of the
websites where the corresponding data can be downloaded.
If the url is denoted by $^{*}$, this means that data were obtained from a
private communication
and they are available
upon request from
the authors of Ref.~\cite{boguna2018}.
}
\begin{tabular}{r|r||r|r|r||r|r|r||r|l}

\multicolumn{2}{c}{} & \multicolumn{3}{c}{Louvain} &
                                                     \multicolumn{3}{c}{Infomap}
  & \multicolumn{2}{c}{}
\\

network & $N$ & $C$ & $Q$ & $\bar{\xi}$ & $C$ & $Q$ & $\bar{\xi}$  & Refs.  & url
\\
\hline
\hline

%\\
\hline
IPv4 Internet & $37,542$
& $31$ & $0.61$ & $0.72$
& $1,625$ & $0.47$ & $0.94$
& \cite{kleineberg2016hidden} & \scriptsize{\url{http://koljakleineberg.wordpress.com/materials}}

\\
\hline
IPv6 Internet & $5,143$
& $19$ & $0.48$ & $0.53$
& $418$ & $0.41$ & $0.86$
& \cite{kleineberg2016hidden} & \scriptsize{\url{http://koljakleineberg.wordpress.com/materials}}

\\
\hline
C. Elegans, layer 1 & $248$
& $9$ & $0.65$ & $0.70$
& $29$ & $0.61$ & $0.83$
& \cite{kleineberg2016hidden, chen2006wiring, de2015muxviz} & \scriptsize{\url{http://koljakleineberg.wordpress.com/materials}}

\\
\hline
C. Elegans, layer 2 & $258$
& $9$ & $0.50$ & $0.82$
& $23$ & $0.46$ & $0.84$
& \cite{kleineberg2016hidden, chen2006wiring, de2015muxviz} & \scriptsize{\url{http://koljakleineberg.wordpress.com/materials}}

\\
\hline
C. Elegans, layer 3 & $278$
& $7$ & $0.44$ & $0.87$
& $11$ & $0.42$ & $0.86$
& \cite{kleineberg2016hidden, chen2006wiring, de2015muxviz} & \scriptsize{\url{http://koljakleineberg.wordpress.com/materials}}

\\
\hline
D. Melanogaster, layer 1 & $752$
& $17$ & $0.64$ & $0.82$
& $70$ & $0.59$ & $0.91$
& \cite{kleineberg2016hidden, stark2006biogrid, de2015structural} & \scriptsize{\url{http://koljakleineberg.wordpress.com/materials}}

\\
\hline
D. Melanogaster, layer 2 & $633$
& $17$ & $0.64$ & $0.72$
& $68$ & $0.60$ & $0.89$
& \cite{kleineberg2016hidden, stark2006biogrid, de2015structural} & \scriptsize{\url{http://koljakleineberg.wordpress.com/materials}}

%%%%%

\\
\hline
arXiv, layer 1 & $1,537$
& $32$ & $0.87$ & $0.78$
& $130$ & $0.81$ & $0.94$
& \cite{kleineberg2016hidden, de2015identifying} & \scriptsize{\url{http://koljakleineberg.wordpress.com/materials}}

\\
\hline
arXiv, layer 2 & $2,121$
& $35$ & $0.86$ & $0.74$
& $190$ & $0.79$ & $0.96$
& \cite{kleineberg2016hidden, de2015identifying} & \scriptsize{\url{http://koljakleineberg.wordpress.com/materials}}

\\
\hline
arXiv, layer 3 & $129$
& $10$ & $0.81$ & $0.88$
& $17$ & $0.78$ & $0.93$
& \cite{kleineberg2016hidden, de2015identifying} & \scriptsize{\url{http://koljakleineberg.wordpress.com/materials}}

%%%%%%

\\
\hline
arXiv, layer 4 & $3,669$
& $46$ & $0.82$ & $0.69$
& $290$ & $0.74$ & $0.91$
& \cite{kleineberg2016hidden, de2015identifying} & \scriptsize{\url{http://koljakleineberg.wordpress.com/materials}}

%%%%%%%%

\\
\hline
arXiv, layer 5 & $608$
& $23$ & $0.85$ & $0.86$
& $61$ & $0.79$ & $0.96$
& \cite{kleineberg2016hidden, de2015identifying} & \scriptsize{\url{http://koljakleineberg.wordpress.com/materials}}

\\
\hline
arXiv, layer 6 & $336$
& $17$ & $0.84$ & $0.96$
& $38$ & $0.80$ & $0.98$
& \cite{kleineberg2016hidden, de2015identifying} & \scriptsize{\url{http://koljakleineberg.wordpress.com/materials}}

\\
\hline
Physician, layer 1 & $106$
& $8$ & $0.51$ & $0.78$
& $13$ & $0.52$ & $0.80$
& \cite{kleineberg2017geometric} & \scriptsize{\url{http://koljakleineberg.wordpress.com/materials}}

\\
\hline
Physician, layer 2 & $113$
& $10$ & $0.56$ & $0.79$
& $14$ & $0.55$ & $0.77$
& \cite{kleineberg2017geometric} & \scriptsize{\url{http://koljakleineberg.wordpress.com/materials}}

\\
\hline
Physician, layer 3 & $110$
& $9$ & $0.60$ & $0.53$
& $18$ & $0.59$ & $0.72$
& \cite{kleineberg2017geometric} & \scriptsize{\url{http://koljakleineberg.wordpress.com/materials}}

\\
\hline
SacchPomb, layer 1 & $751$
& $21$ & $0.79$ & $0.53$
& $86$ & $0.73$ & $0.83$
& \cite{kleineberg2016hidden, stark2006biogrid, de2015structural}  & \scriptsize{\url{http://koljakleineberg.wordpress.com/materials}}

\\
\hline
SacchPomb, layer 2 & $182$
& $13$ & $0.82$ & $0.79$
& $28$ & $0.78$ & $0.91$
& \cite{kleineberg2016hidden, stark2006biogrid, de2015structural}  & \scriptsize{\url{http://koljakleineberg.wordpress.com/materials}}

\\
\hline
SacchPomb, layer 3 & $2,340$
& $25$ & $0.52$ & $0.78$
& $119$ & $0.47$ & $0.88$
& \cite{kleineberg2016hidden, stark2006biogrid, de2015structural}  & \scriptsize{\url{http://koljakleineberg.wordpress.com/materials}}

\\
\hline
SacchPomb, layer 4 & $819$
& $11$ & $0.60$ & $0.69$
& $67$ & $0.56$ & $0.88$
& \cite{kleineberg2016hidden, stark2006biogrid, de2015structural}  & \scriptsize{\url{http://koljakleineberg.wordpress.com/materials}}

\\
\hline
Human brain, layer 1 & $85$
& $5$ & $0.62$ & $0.87$
& $8$ & $0.62$ & $0.92$
& \cite{kleineberg2017geometric}  & \scriptsize{\url{http://koljakleineberg.wordpress.com/materials}}

\\
\hline
Human brain, layer 2 & $78$
& $6$ & $0.55$ & $0.85$
& $8$ & $0.56$ & $0.88$
& \cite{kleineberg2017geometric}  & \scriptsize{\url{http://koljakleineberg.wordpress.com/materials}}

\\
\hline
Rattus, layer 1 & $1,866$
& $32$ & $0.69$ & $0.71$
& $129$ & $0.65$ & $0.87$
& \cite{kleineberg2016hidden, stark2006biogrid, de2015structural}  & \scriptsize{\url{http://koljakleineberg.wordpress.com/materials}}

\\
\hline
Rattus, layer 2 & $529$
& $20$ & $0.85$ & $0.75$
& $61$ & $0.80$ & $0.93$
& \cite{kleineberg2016hidden, stark2006biogrid, de2015structural}  & \scriptsize{\url{http://koljakleineberg.wordpress.com/materials}}

\\
\hline
Air/Train, layer 1 & $69$
& $5$ & $0.34$ & $0.68$
& $6$ & $0.19$ & $0.62$
& \cite{kleineberg2017geometric}  &
                                    \scriptsize{\url{http://koljakleineberg.wordpress.com/materials}}

\\
\hline
Air/Train, layer 2 & $67$
& $6$ & $0.26$ & $0.73$
& $5$ & $0.04$ & $0.41$
& \cite{kleineberg2017geometric}  & \scriptsize{\url{http://koljakleineberg.wordpress.com/materials}}

%%%%%%

\\
\hline
ARK200909 & $24,091$
& $29$ & $0.62$ & $0.77$
& $980$ & $0.53$ & $0.94$
& \cite{papadopoulos2015network_b} & \scriptsize{\url{http://bitbucket.org/dk-lab/2015_code_hypermap}}

\\
\hline
ARK201003 & $26,307$
& $29$ & $0.62$ & $0.71$
& $1,070$ & $0.52$ & $0.94$
& \cite{papadopoulos2015network_b} & \scriptsize{\url{http://bitbucket.org/dk-lab/2015_code_hypermap}}

\\
\hline
ARK201012 & $29,333$
& $28$ & $0.60$ & $0.80$
&$1,171$ & $0.49$ & $0.94$
& \cite{papadopoulos2015network_b} &
\scriptsize{\url{http://bitbucket.org/dk-lab/2015_code_hypermap}}

%%%%%%%%%%%%%

\\
\hline
Enron emails & $33,696$
& $291$ & $0.58$ & $0.66$
& $1,546$ & $0.52$ & $0.82$
& \cite{leskovec2009community, boguna2018} & $^{*}$ %\scriptsize{\url{-}}

\\
\hline
Music chords & $2,476$
& $8$ & $0.29$ & $0.57$
& $6$ & $0.00$ & $0.16$
&  \cite{boguna2018,serra12_music}  & $^{*}$ %\scriptsize{\url{-}}

\\
\hline
OpenFights Air Transp.& $3,397$
& $26$ & $0.65$ & $0.89$
& $167$ & $0.61$ & $0.95$
&  \cite{boguna2018,kunegis13konect_data}  & $^{*}$ %\scriptsize{\url{-}} %\scriptsize{\url{http://konect.uni-koblenz.de/networks/openflights}}

%\\
%\hline
%D. Melanogaster & $1,770$
%& $8$ & $0.39$ & $0.56$
%& $118$ & $0.28$ & $0.72$
%&   \cite{boguna2018} & -\afc{Couldn't find this network in their paper}

\\
\hline
Human Metabolites & $1,436$
& $18$ & $0.67$ & $0.78$
& $101$ & $0.62$ & $0.90$
&   \cite{boguna2018,serrano12_metabolic}  & $^{*}$ %\scriptsize{\url{-}}

\\
\hline
Human HI-II-14 proteome & $4,100$
& $42$ & $0.47$ & $0.54$
& $334$ & $0.43$ & $0.80$
&   \cite{boguna2018,rolland14_proteome}  & $^{*}$ %\scriptsize{\url{-}}

%\\
%\hline
%Words & $7,377$
%& $15$ & $0.24$ & $0.42$
%& $-$ & $-$ & $-$
%&  \cite{boguna2018}  & -

\\
\hline
AS Internet & $23,748$
& $24$ & $0.60$ & $0.78$
& $994$ & $0.52$ & $0.94$
& \cite{boguna2018,Internet_data} & $^{*}$ %\scriptsize{\url{-}}
                                %% \scriptsize{\url{http://www.caida.org/projects/ark}}

%%%%%%%%%%%%%

\\
\hline
AS Oregon Interent, $T = 0.58$ & $6,474$
& $31$ & $0.63$ & $0.66$
& $412$ & $0.54$ & $0.88$
& \cite{leskovec2005graphs} & \scriptsize{\url{http://snap.stanford.edu/data/as.html}}

\\
\hline
Air Transportation, $T=0.14$ & $3,618$
& $36$ & $0.69$ & $0.93$
& $246$ & $0.64$ & $0.97$
& \cite{Airports} & \scriptsize{\url{http://seeslab.info/downloads}}

\\
\hline
P2P, $T = 0.92$ & $6,299$
& $19$ & $0.47$ & $0.77$
& $598$ & $0.41$ & $0.85$
& \cite{p2p_1} & \scriptsize{\url{http://snap.stanford.edu/data/p2p-Gnutella08.html}}

\\
\hline
Euro Roads, $T = 0.28$ & $1,039$
& $25$ & $0.86$ & $0.36$
& $134$ & $0.77$ & $0.71$
& \cite{Euro_roads} & \scriptsize{\url{http://konect.uni-koblenz.de/networks/subelj_euroroad}}

\\
\hline
PSOM, $\langle k \rangle = 5$, $\gamma = 2.1$, $T = 0.1$ & $4,114$
& $40$ & $0.85$ & $0.99$
& $248$ & $0.77$ & $1.00$
& \cite{papadopoulos2012popularity} & \scriptsize{\url{http://www.cut.ac.cy/eecei/staff/f.papadopoulos}}

\\
\hline
PSOM, $\langle k \rangle = 5$, $\gamma = 2.1$, $T = 0.9$ & $4,180$
& $30$ & $0.70$ & $0.75$
& $461$ & $0.58$ & $0.85$
& \cite{papadopoulos2012popularity} & \scriptsize{\url{http://www.cut.ac.cy/eecei/staff/f.papadopoulos}}

\\
\hline
\hline
\end{tabular}

\label{table}

\end{table*}

\subsection{Community structure
and robustness of real-world multiplex networks}

\begin{figure*}[!htb]
\begin{center}
\includegraphics[width=0.8\textwidth]{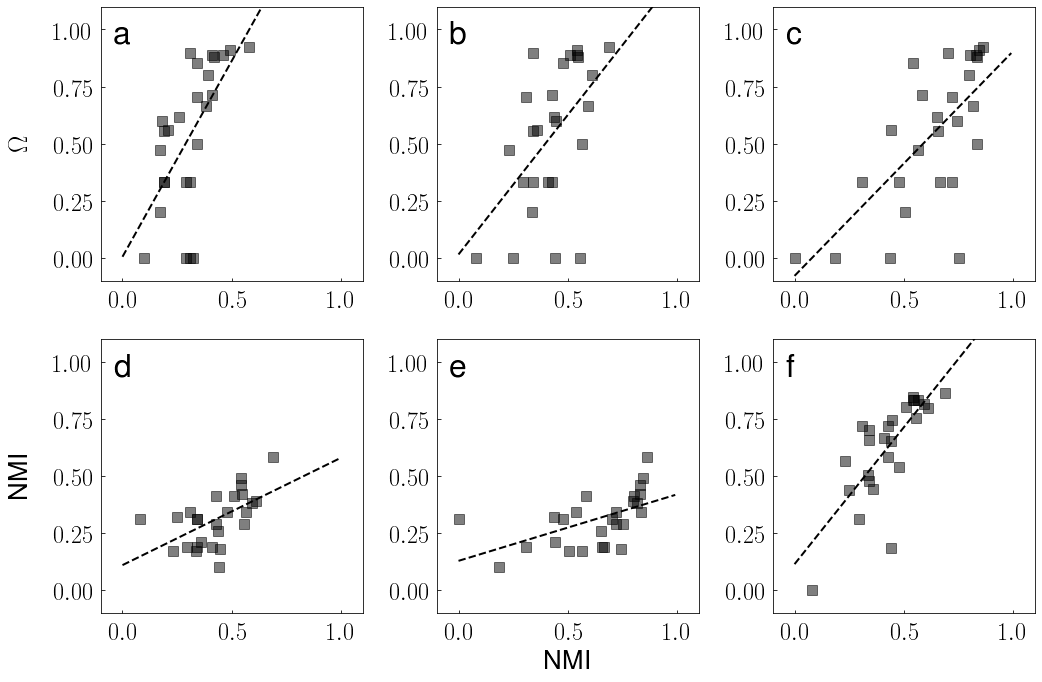}
\end{center}
\caption{Community structure
and robustness of real multiplex networks.
We consider the $26$ multiplex networks analyzed in
  Ref.~\cite{kleineberg2017geometric}. As in
  Ref.~\cite{kleineberg2017geometric}, we rely on the quantity
  $\Omega$ as a proxy to evaluate
  the robustness of a given multiplex network. $\Omega = (\Delta N -
  \Delta N_{rs}) / (\Delta N + \Delta N_{rs})$,  where $\Delta N$ and
  $\Delta N_{rs}$ are respectively the widow sizes of the transitions
  in targeted and random percolation processes on the network.
$\Omega$ values for the
  various networks have been taken from the supplemental material of
Ref.~\cite{kleineberg2017geometric}.
The normalized mutual   information (NMI)
serves to quantify similarity between the
  embedding of the nodes in the two layers. Values of the NMI for
  hyperbolic embedding  have also been taken
from the supplemental material of
Ref.~\cite{kleineberg2017geometric}. We calculated instead NMI values
among the community structures found for the layers of a multiplex using the definition
provided in Ref.~\cite{danon2005comparing}. Communities in each layer are found
using either Louvain or Infomap. (a) As a reference, we reproduced the same plot as in Fig. 4
of Ref.~\cite{kleineberg2017geometric}, where each network represents a
point in the $($NMI, $\Omega)$-plane.  The dashed line is obtained
with simple linear regression. The correlation
coefficient calculated from the data points is $r=0.63$. (b) Same as
in panel a, but for NMI values
calculated using the community structures found by the Louvain
algorithm. Here $r = 0.54$. (c) Same as in panel b, but for NMI values
 calculated using the community structures found by Infomap. We measured
 $r = 0.68$ in this case. (d, e, and f) We compare NMI values obtained using
 the various embedding methods. The various panels represent:
(d) Louvain {\it vs.}  hyperbolic ($r = 0.56$); (e)
 Infomap {\it vs.}  hyperbolic ($r = 0.55$);
(f) Infomap {\it vs.}  Louvain ($r = 0.76$).
}
\label{fig:real}
\end{figure*}

We performed the same
type of analysis as in Ref.~\cite{kleineberg2017geometric} by studying the
relation between system robustness and ``geometric'' correlations
among the network layers in real multiplex networks.  We just
replaced hyperbolic embedding with community structure.
Specifically, given a multiplex network composed of two layers, we
first detect communities in the largest connected component of both
layers independently by using either
Louvain or Infomap.
Correlation between the community structure of the layers
is measured using
the normalized mutual information (NMI) defined in
Ref.~\cite{danon2005comparing}.  As the number of nodes in the layers may be
different, in the computation of the NMI values, we considered only nodes
appearing in both layers. We finally used the
obtained NMI values in the
scatter plots of Fig.~\ref{fig:real}. We find that the
robustness of the various networks can be predicted equally well
by looking at correlations among either hyperbolic coordinates
or community memberships of the nodes
in the two layers (see panels a--c). Further, we find that NMI values
in the various representations are strongly correlated (panels
d--e).

\subsection{Multiplex networks with correlated community structure}

%%%%%%%%
\begin{figure}[!b]
\begin{center}
\includegraphics[width=0.45\textwidth]{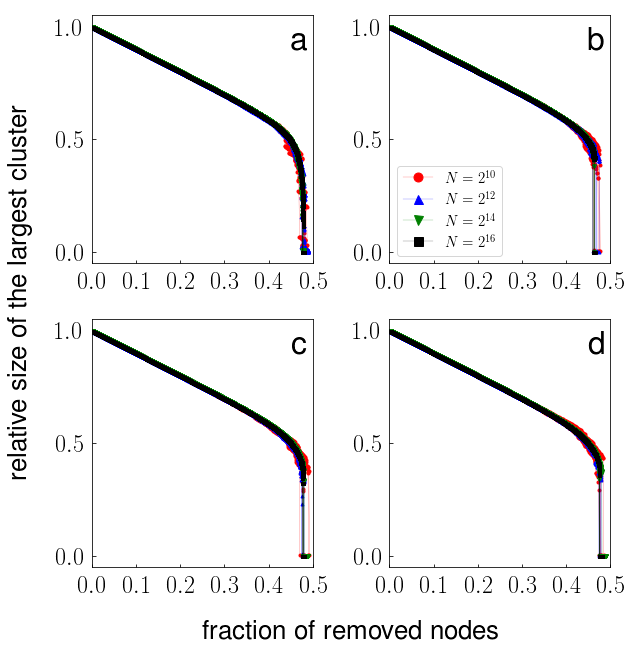}
\end{center}
\caption{Robustness of multiplex networks with correlated community
  structure. We measure the relative size of the largest mutually
  connected cluster
  as a function of the fraction of nodes removed
  from the system.~The
  synthetic
  multiplex graphs
  were
  obtained using the recipe described in the text,  where two
Lancichinetti-Fortunato-Radicchi (LFR)
models %of different
with
 size $N$ are coupled together.
 The LFR models are such that: the average degree is $\langle k
 \rangle = 6$ and
the maximum degree is $k_{max} = 6$, so that degree of all nodes is $k
= 6$; communities have identical size $S = 64$.
For every value of $N$ we show the results for five distinct
instances of the model. (a) LFR graphs are generated with
$\mu = 0.1$. Labels are exchanged only among
nodes within the same clusters. (b) Same as in panel a. However,
relabeling of nodes is allowed among all nodes in the
network. Probabilities of relabeling in panels a and b are such that
the edge overlap among layers is the same for both models.
(c and d) Same as in panels a and b, respectively, but for
LFR graphs constructed using $\mu = 0.3$.
}
\label{fig:2_app}
\end{figure}
%%%%%%%%%

The first step in the creation of a single instance of our multiplex model
consists in generating  a single instance of the Lancichinetti-Fortunato-Radicchi
(LFR) model~\cite{lancichinetti2008benchmark}.
The LFR model is a variant of the degree-corrected stochastic block model. The
model allows to generate single-layer networks with built-in community
structure, where both the degree distribution $P(k)$ and community size
distribution $P(S)$ are power-law functions, i.e., $P(k) \sim
k^{-\gamma}$ and $P(S) \sim S^{-\beta}$. In addition to the exponents
$\gamma$ and $\beta$, in the generation of one
instance of the LFR model, one needs to specify the value of several
parameters, including: average degree $\langle k \rangle$, maximum degree
$k_{max}$, minimum $s_{min}$  and maximum $s_{max}$ community size,
size of the network $N$, and the mixing parameter $\mu$. The mixing parameter
$0 \leq \mu \leq 1$ specifies the fraction of edges that a single
node shares with nodes outside its own community.
This parameter plays a fundamental role to determine how strong the
community structure is. Low values of $\mu$ correspond to a strong
community structure.
As $\mu$ increases, community structure becomes fuzzy.
The maximal value of $\mu$ for which  planted
community structure is exactly recoverable is bounded
by a quantity
calculated in Ref.~\cite{radicchi18dec}. In our simulations, we use
$\mu = 0.1$ to represent a regime of strong community structure, and
$\mu = 0.3$ for regime of loose community structure.
These values have been chosen arbitrarily, thinking to the application
of the model here. For example,
we didn't   use $\mu$ values too close to zero to avoid
the presence of disconnected components.

Once a single instance of the LFR model is generated, we use that
instance to define the topology of both layers of the multiplex.
Node labels of the two layers are initially identical, so that the
adjacency matrices of the
two layers are identical. We then start relabeling nodes of one layer only.
As already mentioned in the main text, we use two different strategies
for relabeling. In the first strategy, we make use of the known
community  structure. In essence, in the relabeling procedure, the label
of every node is exchanged with the one of another
node randomly chosen from the same community.
In the other procedure instead, the constraint on the group
memberships  is not used.  This second variant corresponds to the same model already
considered in Refs.~\cite{radicchi2015percolation,
  osat2017optimal}. In this second variant, we perform a number of
label swaps such that the value of the edge overlap among the two layers is
comparable  with the one obtained in the first variant of the model.
Both variants of the multiplex model essentially lead to very small
values of  edge overlap and degree-degree correlation
between layers. The first variant, however, preserves perfect
correlation  between the community structure of the two layers, while
the second variant destroys it completely.

The robustness of
single instances of the multiplex model described above are then
studied as in Ref.~\cite{kleineberg2017geometric}. Every node $i$ in
the network has associated the score $K_i = \max {(k_i^{(1)},
  k_I^{(2)})}$, with $k_i^{(x)}$ the degree of node $i$ in layer $x$.
Nodes are then ranked in descending order according to this score,
with ties randomly broken. The top node is removed from the
network. After every removal,
the score is $K_i$ of every node $i$ still in the system is
recomputed. Further, the relative
size of the mutually connected giant component  is
evaluated to construct a percolation phase diagram~\cite{buldyrev2010catastrophic}.

We considered various sets of parameters for the generation of the LFR
model. All of them provide the same type of message. When the community
structure is strong (i.e., small $\mu$ values), the model with
correlated community structure undergoes a smooth percolation
transition.  If correlation in community structure is
destroyed,  the transition becomes abrupt. If the community structure is not
strong (i.e., large $\mu$ values), then both relabeling schemes lead to
an abrupt transition. The result is valid also for LFR models with
homogenous degree distribution (see Figure~\ref{fig:2_app}).

\subsection{Greedy routing}

%%%%%%%%
\begin{figure}[!t]
\begin{center}
\includegraphics[width=0.48\textwidth]{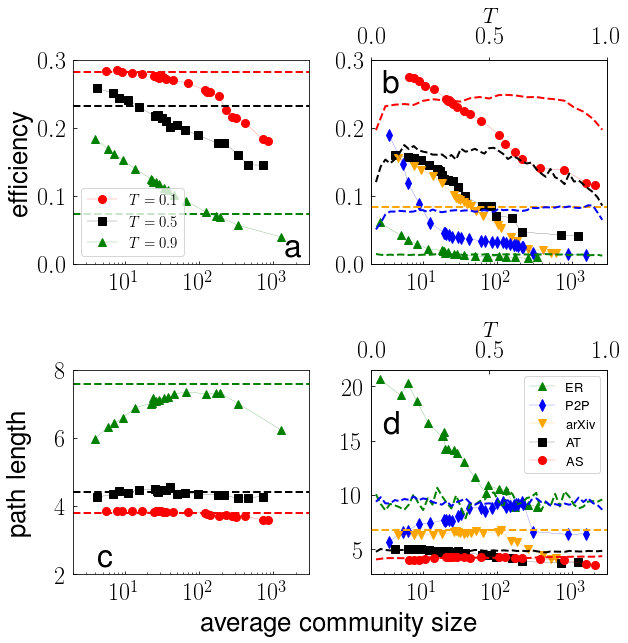}
\end{center}
\caption{Same analysis as in Figure~3 of the main text.
Performance is measured in terms of efficiency (panels a and b), and
the average path length of successfully delivered packets (panels c and d).
}
\label{fig:3_app}
\end{figure}
%%%%%%%%%

As already considered in Refs.~\cite{boguna2009navigability,
boguna2010sustaining}, we immagine that a packet is traveling
from the source node $s$ to the
target node $t$ in a network with $N$ nodes
and adjacency matrix $A$. The packet moves on edges of the network,
performing a single hop at each stage of the dynamics. Greedy routing
relies on a definition of ``distance'' between pairs of nodes in the network.
At every stage $r$ of the dynamics towards the target node $t$, a packet
sitting on node $p_r = i$ choose to move to the node $j^{\,(i)}_{(best)}$ defined in
Eq.~(4) of the main text. In essence,  $j^{\,(i)}_{(best)}$ is
the neighbor of node $i$ that has minimal distance to the target node
$t$. %Eventual ties are randomly broken.
In our numerical simulations,
we avoid  immediate backtracking walks of the
packet, therefore node $j^{\,(i)}_{(best)} = p_{r+1} \neq p_{r-1}$, i.e.,
cannot be equal to the node visited before node $i$; this condition improves significantly (not shown) the performance of both methods considered in this paper.
The packet continues to travel until one of these two conditions is
met: (i) the packet arrives at destination  after $R$ steps, i.e., $p_{R}
= t$; (ii) the packet visits twice the same node, i.e., $p_{r} = p_{v}$,
with $v<r$. Condition (i) corresponds to success. Condition (ii)
represents failure and the packet is discarded.
To evaluate performance of the routing protocol, we use at least $B = 10,000$ numerical simulations. In each simulation,
source $s$ and target $t$ nodes are randomly chosen among the
nodes in the giant connected component of the network.
We quantify three different metrics of performance:
\begin{itemize}
\item[1)] The success rate $z$
, i.e., the fraction of packets
  correctly delivered. This is a metric of performance introduced
in Ref.~\cite{boguna2009navigability}. Results for this metric are
presented in Figure~3 of the main text.

\item[2)] The average value of $\langle R \rangle$, i.e., the average length of the paths for
successfully delivered packets. This metric of performance was also introduced
in Ref.~\cite{boguna2009navigability}. Results for this metric are
presented in Figures~\ref{fig:3_app}c and d.

\item[3)] Efficiency $\eta=z  \,\langle 1/R \rangle$,
  where $\langle 1/R \rangle$ represents the mean value of the inverse
  of the
path length obtained for each of the successfully delivered packets. This definition of $\eta$ is based on a metric of performance introduced
in Ref.~\cite{muscoloni2017machine}. Results for this metric are
presented in Figures~\ref{fig:3_app}a and b.

\end{itemize}
It is worth noting that the efficiency measure (which is a balance between success rate and path length) shows similar results as those of the success rate (Figure~\ref{fig:3_app}a and b); this is because for almost all the networks of Figure~\ref{fig:3_app}, the path length does not change remarkably as the mean community size or the temperature is altered (Figures~\ref{fig:3_app}c and d). Thus, the success rate results (investigated in Figure~3 of the main text) are sufficient to assess the performance of the two routing methods investigated in this paper.

%%%%%%%%
\begin{figure}[!b]
\begin{center}
\includegraphics[width=0.5\textwidth]{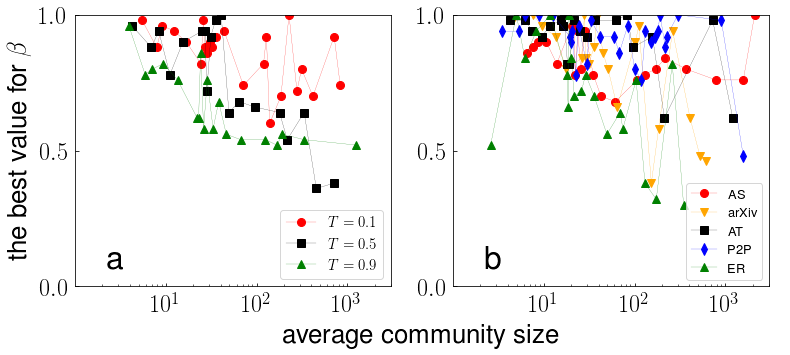}
\end{center}
\caption{Same analysis as in Figure~3 of the main text. For each modular structure the $\beta$ value for which we obtained the highest success rate is reported.
}
\label{fig:3_app_beta}
\end{figure}
%%%%%%%%%
In the standard application
of network hyperbolic embedding, the distance between
pairs of nodes is
given by their distance in the hyperbolic
space~\cite{boguna2009navigability,
boguna2010sustaining}. In our community-based routing protocol, we substituted the distance
in the hyperbolic space with the analogous quantity
based on the {\it a priori} given community structure
of the graph. Specifically, we define the weight between the
connected modules $g$ and $q$ as
\begin{equation}
w_{g,q} = 1 - \ln {\rho_{g,q}} \qquad \textrm{ , if } \rho_{g,q} > 0 \; ,
\label{eq:comm_dist}
\end{equation}
where
\begin{equation}
\rho_{g,q} = \frac{\sum_{i > j} A_{i,j} \delta_{\sigma_i, g}
\delta_{\sigma_j, q} } {\sum_{i > j} A_{i,j} \delta_{\sigma_i, g}} \; .
\label{eq:density}
\end{equation}
In the above equation,
$\delta_{x,y} = 1$, if $x=y$, while $\delta_{x,y} = 0$, otherwise;
$A_{i,j} = A_{j,i} = 1$ if nodes $i$ and $j$ are connected, while
$A_{i,j} = A_{j,i} = 0$, otherwise; $\sigma_i$ is the group membership
of node $i$ according to the given community structure.
Eq.~(\ref{eq:density})
is the ratio between the total number of edges
shared between nodes within communities $g$ and $q$, and the total
degree of nodes in community $g$.
$\rho_{g,q}$ can be also interpreted as the
 probability that following a random edge of a random node in module
 $g$ we reach a node in module $q$.
We consider each community as a supernode,
and the network as a supernetwork composed of supernodes connected
with weighted superedges.
The weight of the superedge between supernodes $g$ and $q$ is defined
in Eq.~(\ref{eq:comm_dist}).
Then, we find the length of the shortest paths between every pair of
supernodes. This operation relies on the algorithm by
Johnson~\cite{johnson1977efficient}.
The output is a full matrix $D$ that includes the
distances between every pair of modules. The generic element $D_{g,q}$
of this matrix contains a sum of weights defined in
Eq.~(\ref{eq:comm_dist}), which is basically equivalent to a sort
of expected path length between communities $g$ and $q$, under the
hypothesis that  connections
were generated according to the
stochastic block
model~\cite{karrer2011stochastic}. Given that we are at node $i$
at stage $r$ of the trajectory of the packet, the ``distance'' between a
neighbor $j$ of node $i$ and the target $t$ is finally defined as
\begin{align}
d_{j, t} =&\, \beta D_{\sigma_j, \sigma_t}  \nonumber \\&+ (1 - \beta) \left\{ \left[1 - \log {\left( k_j \, \rho_{\sigma_i,
      \sigma_j} \right)} \right]  -  \left[ 1 - \log {\left(  \rho_{\sigma_i,
      \sigma_j} \right)} \right]   \right\} \\
      =&\, \beta D_{\sigma_j, \sigma_t} - (1 - \beta) \ln { k_j }
\end{align}
where $k_j$ is the degree of node $j$, and $0 \leq \beta \leq 1$.
The previous expression defines a measure of
``distance" between node $j$ and module $\sigma_t$. This is computed
as a distance between modules $\sigma_j$ and $\sigma_t$, but corrected
for the fact that we are aware of the degree of node $j$. This
definition of distance is motivated by the degree-corrected stochastic
block model in which the
%probability that one randomly chosen edge from community $\sigma_i$ is attached to node $j$
probability that following a randomly chosen edge from community $\sigma_j$ we reach a node in community $\sigma_{j'}$ is proportional to $k_j \, \rho_{\sigma_j,\sigma_{j'}}$.
Note that we are
aware also of the degrees of nodes $i$ and $t$, but this information is not
helpful in the protocol. %$k_j\,\rho_{\sigma_i,\sigma_j}$.
The factor $\beta$ serves to weight the importance
    of the community structure {\it vs.} the degree of the individual
    nodes in the definition of distance. This factor can be tuned
    appropriately to optimize the success rate of the greedy routing
    protocol. Optimal values used in our simulations are displayed in
    Figure~\ref{fig:3_app_beta}. As Figure~\ref{fig:3_app_beta} illustrates, the most optimum value of $\beta$ depends on the network structure and also on the considered modular structure;
    more specifically, $\beta$ is more likely to be close to 1 for networks with lower temperatures (or effectively those with higher clustering coefficients) and for modular structures with smaller mean community sizes.
     %The former expression reduces to
%Eq.~(\ref{eq:distance}) of the main text.

%%%%%%%%%%%%
%%%%%%%%%%%%
%%%%%%%%%%%%

%\bibliography{all_bibs}


\end{document}